\documentclass[aps,prd,floats]{revtex4}
\input epsf

\def\rn{}
\def\nn#1 #2{#2. #1}				
\def\nnn#1 #2 #3{#2. #3. #1}			
\def\nnnn#1 #2 #3 #4{#2. #3. #4 #1}		
\def\nnnnn#1 #2 #3 #4 #5{#2. #3. #4 #5. #1}	
\def\dualand{ and\hbox{ }}				
				
\def\rf#1;#2;#3;#4;#5 {{\frenchspacing\par\rn#1, #3 {\bf #4}, #5 (#2). \par}}
\def\rg#1;#2;#3;#4;#5;#6 {{\frenchspacing\par\rn#1, #3 {\bf #4}, #5 (#2). \par}}
\def\rfbook#1;#2;#3;#4;#5 {{\frenchspacing\par\rn#1, {\it #3} (#5, #4, #2).\par}}
\def\rfprep#1;#2;#3 {{\par\frenchspacing\rn#1, #3 (#2).\par}}
\def\rfproc#1;#2;#3;#4;#5;#6 {{\frenchspacing\par\rn#1 #2, in {\it #3}, ed. #4 (#5: #6)\par}}
\def\rfprocp#1;#2;#3;#4;#5;#6;#7 {{\frenchspacing\par\rn#1 #2, in {\it #3}, ed. #4 (#5: #6), p#7\par}}

\newcommand{\beq}{\begin{equation}}
\newcommand{\eeq}{\end{equation}}

\def\lap{\lower.5ex\hbox{$\; \buildrel < \over \sim \;$}}
\def\gap{\lower.5ex\hbox{$\; \buildrel > \over \sim \;$}}

\def\L{\Lambda}

\def\ba{\begin{eqnarray}}
\def\ea{\end{eqnarray}}
\def\be{\begin{eqnarray}}
\def\ee{\end{eqnarray}}

\def\erfc{{\rm erfc}\,}

\begin{document}

\title{Anthropic predictions for vacuum energy and neutrino masses in the light of WMAP-3}

\author{Levon Pogosian{$^1$} and Alexander Vilenkin{$^2$}}

\affiliation{ $^1$ Department of Physics, Simon Fraser University, \\ 8888 University Drive,
Burnaby, BC, V5A 1S6, Canada \\
$^2$ Institute of Cosmology, Department of Physics and Astronomy,\\ 
Tufts University, Medford, MA 02155, USA}

\begin{abstract}

Anthropic probability distributions for the cosmological constant
$\Lambda$ and for the sum of neutrino masses $m_\nu$ are updated using
the WMAP-3 data release. The new distribution for $\Lambda$ is in a
better agreement with observation than the earlier one.  The
typicality of the observed value, defined as the combined probability
of all values less likely than the observed, is no less than 22\%.  We
discuss the dependence of our results on the simplifying assumptions
used in deriving the distribution for $\Lambda$ and show that the
agreement of the anthropic prediction with the data is rather robust.
The distribution for $m_\nu$ is peaked at $m_\nu\sim 1$~eV, suggesting
degenerate neutrino masses, but a hierarchical mass pattern is still
marginally allowed at a $2\sigma$ level.

\end{abstract}

\maketitle

\section{Introduction}
\label{sec:intro}

The parameters we call constants of Nature may in fact be stochastic
variables taking different values in different parts of the universe.
The observed values are then determined by chance
and by anthropic selection. Recent developments in string theory
\cite{Bousso,Susskind03}, combined with the ideas of eternal inflation
\cite{AV83,Linde86}, have led to the "landscape" paradigm, which
provides a theoretical basis for this picture. And the successful
prediction of a nonzero cosmological constant $\Lambda$
\cite{Weinberg87,Linde87,AV95,Efstathiou95,Weinberg97,MSW} may be our first evidence
for the existence of a huge multiverse beyond our horizon. (For an up
to date discussion of these ideas, see \cite{Carr}.)

Anthropic predictions are necessarily statistical in nature. Even
though the observed value of $\Lambda$ is within the expected range,
there is a lingering doubt that this may be no more than a lucky
guess.  To further test the theory, we should try to extract
predictions for variables other than $\Lambda$. Some attempts in this
direction have already been made. Predictions for the neutrino masses
$m_\nu$ have been discussed in \cite{TVP03,PVT04}, and a postdiction for
the (axionic) dark matter mass per baryon, $\xi$, has been given in
\cite{Wilczek}. 

A reliable calculation of anthropic probability distributions is a
very challenging task. All calculations performed so far have relied
on a number of simplifying assumptions. Moreover, the form of the
distributions for $\Lambda$ and $m_\nu$ depends on the values of the
cosmological parameters: the Hubble parameter $h$, the tilt of the
perturbation spectrum $n_s$, and particularly on $\sigma_8$ -- the
amplitude of linear density perturbations on the scale of $8h^{-1}$
Mpc.  Hence, the agreement (or disagreement) of the predictions with
the data should be regarded as tentative.  With improved understanding
or improved measurements, it may either get better or worse.

\begin{figure}[tb]
\centerline{\epsfxsize=9.0cm\epsffile{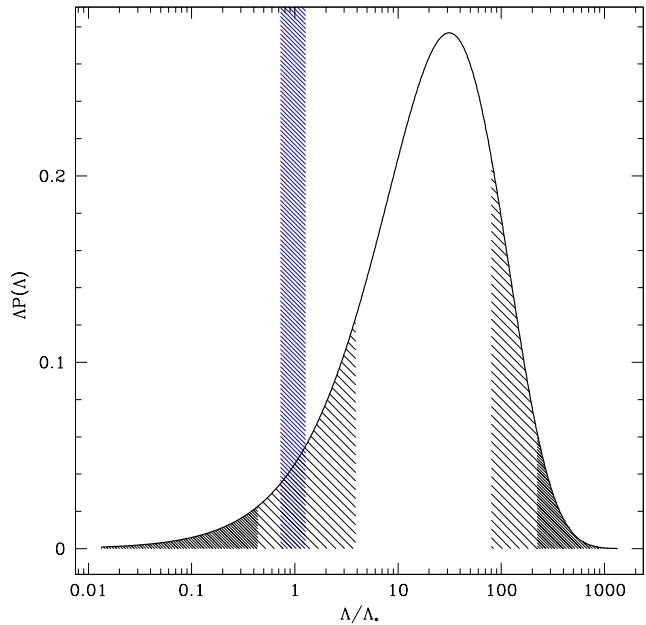}}
\caption{\label{prho05}\footnotesize
The probability distribution for $\Lambda$ based on the 
flat $\Lambda$CDM best fit parameters from WMAP-1. 
$\Lambda_*$ is the observed value whose $1\sigma$ uncertainty
is represented by the blue shaded vertical stripe.  
}
\end{figure}

In Fig.~\ref{prho05} we show the distribution $P(\Lambda)$ circa
2005. Its calculation has been discussed in
\cite{AV95,Efstathiou95,AV96,Weinberg97,MSW,GV03,GLV03}. The parameter
values used in the figure were based primarily on the first year of
WMAP data. The agreement between the theory and observation can be
characterized by the ``typicality'' of the observed value $\Lambda_*$,
${\cal T}(\Lambda_*)$ \cite{Page}. (The definition of typicality will
be discussed in Section II.) For the distribution in Fig.1, ${\cal
T}(\Lambda_*)\approx 10$\%. For a statistical model of this sort, this
can be regarded a good agreement.\footnote{A better agreement,
corresponding to a $31$\% typicality, was reported in
\cite{Weinberg05} based on a larger effective galaxy mass scale; see
the discussion at the end of Section III.}

With the release of the 3-year WMAP data, some of the parameters (such
as $n_s$ and $h$) are now known with a better accuracy. Among
implications was a decrease in the value\footnote{Inferred from the
WMAP data alone under the assumption of the flat power law
$\Lambda$CDM model} of $\sigma_8$ by more than 10\%, from $0.9\pm
0.1$\cite{wmap1} to $0.76 \pm 0.05$\cite{wmap3}.  Combining other
datasets with the WMAP data can change the preferred values of the
parameters. For example, adding all available information, including
the SDSS Lyman-$\alpha$ forest data \cite{mcdonald05}, was reported to
increase the value of $\sigma_8$ to $0.85 \pm 0.02$
\cite{seljak06}. On the other hand, combining the WMAP data with the
distribution of SDSS luminous red galaxies yields $\sigma_8=0.76 \pm
0.03$
\cite{tegmark_lrg} in good agreement with results of the WMAP team. 
Here, as in our original analysis, we will make use of the WMAP data
alone, which is most likely to be free of systematics and, for
the flat power law $\Lambda$CDM model, provides sufficiently accurate
measurements of the spectral index and other cosmological parameters
that determine $\sigma_8$.

The main goal of the present paper is to update the distributions for
$\Lambda$ and $m_\nu$ using the WMAP-3 data. We shall see that both
distributions are shifted toward smaller values of $\Lambda$ and
$m_\nu$, respectively. The agreement between theory and observations
is improved as a result. We shall also discuss the dependence of the
results on the simplifying assumptions made in the calculation of
probabilities.

In the next section, we discuss the concept of typicality which will
be used as a quantitative measure of the agreement of theoretical
distributions with the data.  The case where $\Lambda$ is the only
variable is considered in Section \ref{sec:plambda}. The effects of
other variable parameters, such as the amplitude of the primordial
density perturbations $Q$ and the dark matter mass per photon $\xi$,
and the dependence on the assumed value of the galactic mass $M_G$,
are discussed in \ref{sec:mass}. Variable $m_\nu$ and joint variation
of $m_\nu$ and $\Lambda$ is analyzed in Section \ref{sec:nu}. We
finish with some concluding remarks in Section \ref{sec:conclude}.

\section{Typicality}

\begin{figure}[tb]
\centerline{\epsfxsize=9.0cm\epsffile{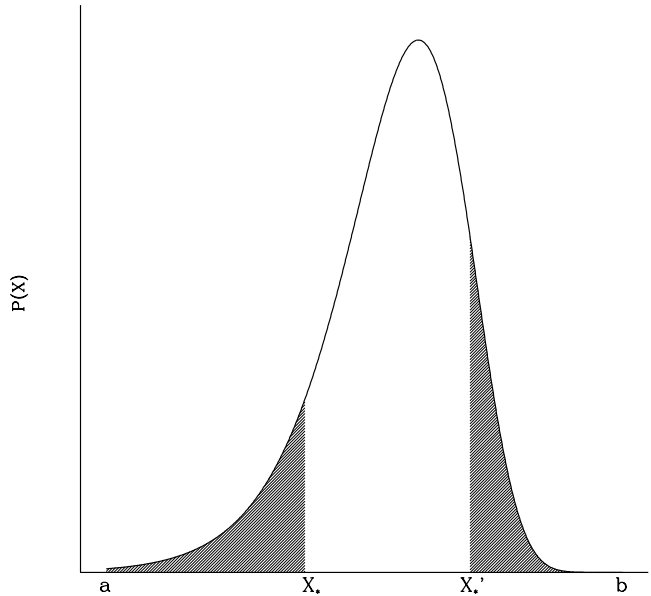}}
\caption{\label{fig:typ}\footnotesize
Values ``less probable'' than $X_*$ are located in the two shaded
regions, which have equal measure in the distribution $P(X)$.
}
\end{figure}

Consider a variable $X$ defined in the interval $a<X<b$ and
characterized by a probability distribution $P(X)$.  The value $X =
X_*$ has typicality ${\cal T}$ if ${\cal T}$\% of the distribution are
``less probable'' than $X_*$. Page \cite{Page} assumes that the least
probable values are at the tails of the distribution (see
Fig.~\ref{fig:typ}). The motivation is that such values are fine-tuned
to be near the special points $a$ and $b$, which are the endpoints of
the interval.

Suppose for definiteness that $X_*$ is closer to $a$ than to $b$ in
terms of the measure $P(X)$, so that 
\beq
P_*\equiv \int_a^{X_*}P(X)dX <{1\over{2}}.
\eeq
Let us also define ${X_*}'$, which is equally close to $b$,
\beq
\int_{{X_*}'}^b P(X)dX = P_*.
\eeq
The values in the intervals $a<X<X_*$ and ${X_*}'<X<b$ are regarded as
less likely than $X_*$. The typicality is then given by
\beq
{\cal T}(X_*)=2P_*.
\label{typicality1}
\eeq

In the case of the cosmological constant, apart from the vicinity of
the endpoints at $\Lambda=\pm\infty$, there is an additional range of
unlikely values near zero.  $\Lambda = 0$ is a special value, which
defines the boundary between eternal expansion and recollapse. Having
$\Lambda$ accidentally fine-tuned very close to this point is
unlikely, and we want this reflected in the typicality. We shall
define $P_*$ as the ``distance'', in terms of the measure $P(\L)$,
from the observed value $\L_*$ to the nearest special point
$(\L=0,\pm\infty)$. It will also be convenient to define the
probabilities $P_\pm$ for positive and negative $\L$,
\beq
P_+=\int_0^\infty P(\L)d\L,
\eeq
\beq
P_-=\int_{-\infty}^0 P(\L)d\L,
\eeq
$P_+ +P_- =1$. A straightforward generalization of Page's definition
is to define typicality as the measure of all points whose distance to
the nearest special point is less than $P_*$.

It is clear from Fig.~\ref{prho05} that the special point closest to
the observed value $\L_*$ is $\L=0$.  Hence,
\beq
P_*=\int_0^{\L_*}P(\L)d\L <P_+/2.
\label{P*P+}
\eeq
If in addition $P_*<P_-/2$, then there are four intervals of $\L$ less
likely than $\L_*$, each having measure $P_*$, and thus
\beq
{\cal T}(\L_*)=4P_*.
\label{typicality2}
\eeq
Alternatively, if $P_*>P_-/2$, we have\footnote{It is instructive to
compare Eqs.~(\ref{typicality2}),(\ref{typicality3}) with Page's
original definition, which involves only two special points at
$\L=\pm\infty$.  It gives ${\cal T}_{Page}(\L_*)=2(P_-+P_*)$ for
$P_-+P_*<P_+$ and ${\cal T}_{Page}(\L_*)=2P_+$ for $P_-+P_*>P_+$.  It
can be easily verified that in both cases Page's definition yields a
greater typicality than ours, ${\cal T}_{Page}(\L_*)>{\cal T}(\L_*).$}
\beq
{\cal T}(\L_*)=P_-+2P_*.
\label{typicality3}
\eeq

Most of the calculations of the distribution $P(\Lambda)$ have been
performed for $\Lambda >0$. For negative $\Lambda$, the universe
recollapses at $t_\Lambda = (6\pi G|\L|)^{-1/2}$. To find the distribution in this case,
we need to know the probability for a civilization to evolve during
this time interval, which is of course very uncertain. It is very
likely, however, that positive $\L$ are more probable than negative $\L$
\cite{Weinberg05}, 
\beq
P_+>P_-.
\label{P+P-}
\eeq 
Note, for example, that for $\Lambda = -20\Lambda_*$ the universe
would recollapse in 3 billion years. Star formation in such a universe
will not begin before the onset of the recollapse, and the remaining
1.5 billion years are hardly sufficient to disperse heavy elements in
supernova explosions, form planetary systems and evolve intelligent
life. On the other hand, the 2$\sigma$ range for positive $\Lambda$
extends to $80 \Lambda_*$. The inequality (\ref{P+P-}) can be used to
derive a lower bound on ${\cal T}(\L_*)$.

Let ${\cal T}_+(\L_*)$ be the typicality calculated in the interval
$0<\L<\infty$ with $P(\L)$ normalized in that interval,
\beq
{\cal T}_+(\L_*)=2P_*/P_+.
\label{T+}
\eeq
Now, using Eq.~(\ref{P+P-}) it is easily verified that
\beq
{\cal T}(\L_*)>{\cal T}_+(\L_*).
\label{TT*}
\eeq

In the rest of the paper we shall consider only positive $\Lambda$ and
the corresponding typicality ${\cal T}_+$. According to (\ref{TT*}),
it can be regarded as a lower bound for the full typicality ${\cal
T}$.

\section{The probability distribution for $\L$}
\label{sec:plambda}

The method used in this paper follows closely that in
\cite{MSW,GV03,TVP03,PVT04}. Here we only reproduce the essential details
and refer the reader to \cite{PVT04} for a comprehensive description.
Consider a model in which $\L$ is allowed to vary from one part of the
Universe to another. We define the distribution ${\cal P}(\L) d\L$ as
the probability for a randomly picked observer to measure $\L$ in
the interval $d\L$.  This distribution can be represented as
\beq 
P(\L)=P_{prior}(\L)f_{selec}(\L),
\label{PPf}
\eeq
where the prior probability $P_{prior}(\L)$ is determined by the
geography of the landscape and by the dynamics of eternal inflation,
and $f_{selec}(\L)$ accounts for anthropic selection effects. 

The standard argument \cite{AV96,Weinberg97} 
suggests that the prior probability is well approximated by
\beq
P_{prior}\approx {\rm const},
\label{flatLambda}
\eeq
because the anthropic range where $f_{selec}(\L)$ is appreciably
different from zero is much narrower than the full range of $\L$. We
emphasize that this is just a heuristic argument. The conditions for
its validity have been discussed both in scalar field models, where
$\L$ is a continuous parameter \cite{GV00,Weinberg00,GV01,GV03}, and in
``discretuum'' models with a large set of metastable vacua 
\cite{Delia1,Delia2}. Future work will show whether or not these
conditions are satisfied in the string theory landscape. Here we shall
assume that Eq.(\ref{flatLambda}) is valid.

It has been the standard practice to identify the selection factor in
(\ref{PPf}) with the asymptotic fraction of baryonic matter,
$F(M>M_G,\L)$, which clusters into objects of mass greater than the
characteristic galactic mass $M_G\sim 10^{12}M_\odot$:
\beq
f_{selec}(\L)\propto F(M>M_G,\L).
\label{fselec}
\eeq
The idea here is that there is a certain average number of stars per
unit baryonic mass and a certain number of observers per star, and
that these numbers are not strongly affected by the value of $\L$. In
the present Section we shall adopt this standard approach; its
validity will be analyzed in Section III.

The fraction of collapsed matter $F(M>M_G,\L)$ can be approximated
using the Press-Schechter (PS) formalism \cite{PressSchechter}. This leads to
\ba
F(M>M_G,\L) &\propto& 
\erfc\left[{X\over\sqrt{2}}\right],
\label{nG}
\ea
where
\beq
X=\delta_c/{\sigma_\infty(\L)} \ ,
\label{X}
\eeq
$\delta_c\approx 1.63$ and $\sigma_\infty$ is the variance of 
the Gaussian density fluctuation field in the asymptotic future on the 
galactic scale $M_G$. The latter quantity can be written as \cite{PVT04}
\ba
\sigma_\infty(\L)= {\sigma_G^*}
{D^{\infty}(\L) \over {D}(\L_*,x_*)} \ .
\label{X2}
\ea 
Here, ${\sigma}_G^*$ is the current density contrast on the galactic
scale $M_G$ inferred from the large-scale CMB data, $x_*$ is our local
value of $\Omega_\L/\Omega_M$, $\L_*$ is the local value of vacuum
energy density, $D(\L_*,x_*)$ is the local linear growth factor and
$D^{\infty}(\L) \equiv D(\L,x=\infty)$. 

N-body simulations indicate that the PS model overestimates the
abandance of ``typical'' halos, while underestimating that of more
massive structures \cite{jenkins01}. The Sheth-Tormen model
\cite{ST99} was shown to fit the simulations better. We have checked
that replacing the PS formula with that of \cite{ST99} does not
significantly affect our results and use the PS model through out this
work.

To evaluate ${\sigma}_G^*$ we first find the length scale
$R(M_G)$ corresponding to the mass scale $M_G$ using 
\ba
R(M_G) &=& \left( {3M_G \over 4\pi \rho_0} \right)^{1/3} \nonumber \\
&\approx& 0.951 \ \left( {h \over \Omega_M}\right)^{1/3}
\left( {M_G \over 10^{12}M_{\odot}} \right)^{1/3} {\rm h^{-1} Mpc} \ ,
\ea
where $\rho_0 \approx 1.88 \cdot 10^{-26} \ {\Omega_M h^2} \ {\rm
kg/m}^3$ is the mean cosmic density of nonrelativistic matter. For
$h=0.73$ and $\Omega_M =0.234$ this gives 
\be
R(M_G) \approx 1.4 \ {\rm h^{-1} Mpc} \left( {M_G \over 10^{12} M_{\odot}} 
 \right)^{1/3}\ .
\ee
The corresponding
linearized density contrast ${\sigma}_G^*$ found using WMAP-3's best fit power 
law model \cite{wmap3} for $M=10^{12} M_{\odot}$ is
\be
{\sigma}_G^* \equiv \sigma[R(M_G)]= \left[ {1 \over 2\pi^2} \int_0^{\infty} P(k) W^2(kR) k^2 dk \right]^{1/2} \approx 1.87 \pm 0.13 \ .
\label{sigmarg}
\ee
In the above, $P(k)$ is today's linear matter power spectrum and $W(kR)$ is the window function.
We work with the ``top-hat'' form for $W(kR)$, also used by the WMAP team.
This value is smaller than ${\sigma}_G^* \approx 2.41 \pm 0.26$ inferred 
from the WMAP-1 data and used in \cite{TVP03,PVT04}.

The growth factor $D(\Lambda,x)$ in a universe containing only $\L$ and pressureless matter can be written as \cite{heath77,MSW}
\be
D(\Lambda,x)=1+{3 \over 2} {x_{eq}}^{-1/3} G(x) \ ,
\ee
where
\be
G(x)={5 \over 6} \left[1+x \over x \right]^{1/2} \int_0^x {dw \over w^{1/6} \ (1+w)^{3/2}} \ . 
\ee
Using ${3 \over 2} {x_{eq}}^{-1/3} G(x) \gg 1$, the variable $X$ defined in Eq.~(\ref{X}) can be written as
\beq
X={\delta_c \over \sigma_G^*} \left(\L \over \L_* \right)^{1/3} {G(x_*) \over G(\infty)} \ .
\eeq

The variance $\sigma_G^*$ is  proportional to the amplitude of primordial fluctuations $Q$. The dependence on Q can be explicitly introduced by writing
\beq
\sigma_G=\sigma_G^* {Q \over Q_*} \ ,
\eeq
where $Q_*$ is the observed value of $Q$. We follow the WMAP-3 team's conventions \cite{wmap3} and define $Q\equiv \Delta_k (k=0.002 {\rm h/Mpc})$ with the best fit value of $Q_*=4.9 \times 10^{-5}$. Then $X$ can be written as
\beq
X=A \left(\L \over \L_* \right)^{1/3}  {1 \over Q} \ ,
\label{X1}
\eeq
with
\beq
A\approx 3.2 \times 10^{-5} \left(M_G \over 10^{12} M_\odot \right)^{0.1} \ ,
\label{A1}
\eeq
where the mass dependence of $A$ comes from that of $\sigma_G^*$ in (\ref{sigmarg}) and the approximation is valid for values of $M_G$ within a few orders of magnitude of $10^{12} M_\odot$.

In summary, the distribution for $\L$ is 
\beq
P(\L)d\L \propto \erfc\left[{X\over{\sqrt{2}}}\right]d\L,
\label{erfc}
\eeq
where $X$ is given by Eq.~(\ref{X1}) with $A$ from Eq.~(\ref{A1}).

The degree of clustering that we observe today 
is often 
parameterized by the variance of the linear density contrast field on the
$8 h^{-1}$Mpc scale, $\sigma_8$. The latest observations,
as mentioned in the introduction, indicate a lower value of $\sigma_8$, 
and correspondingly a lower value of $\sigma_G$. A decrease in
$\sigma_G$ at a fixed $\L$ leads to a suppression of galaxy formation,
except for very small values of $\L$, such that $\L$-domination occurs
when clustering on the galactic scale is essentially complete. As a
result the peak of the probability 
distribution is shifted towards lower values, leading to
a better agreement with the observed $\L$.

\begin{figure}[tb]
\centerline{\epsfxsize=9.0cm\epsffile{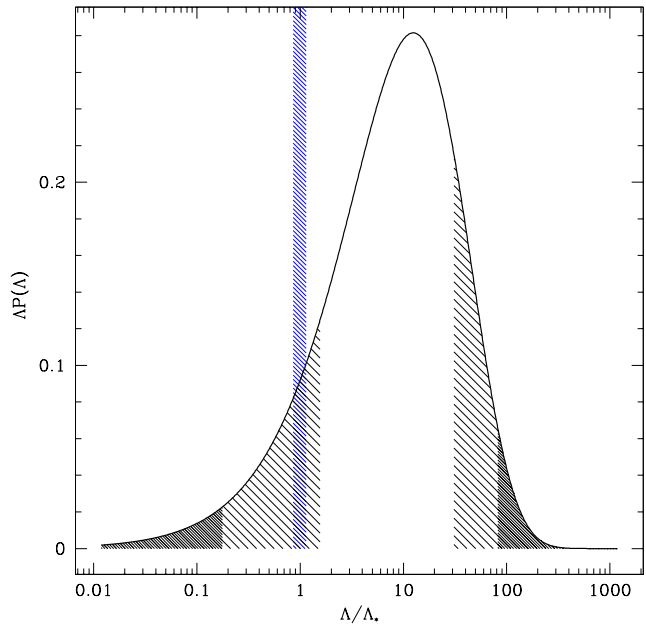}}
\caption{\label{prho06}\footnotesize
The probability distribution for $\Lambda$ obtained as in Fig.~\ref{prho05}
but using parameters derived from the WMAP-3 data. The observed value, 
$\Lambda_*$, has a typicality of $22$\%.}
\end{figure}

The updated distribution is shown in Figure~\ref{prho06} and corresponds to
a typicality of ${\cal T}_+ = 22$\% for our local value of $\L$. 

In \cite{Weinberg05} Weinberg reports a probability of $15.6$\% for
the observed value of $\L$, implying a typicality ${\cal T}_+ = 31$\%.
(His probability includes only values between $0$ and $\L_*$; it is
half of what we call ${\cal T}_+$.) This result is based on the WMAP-1
data and is higher because of the use of the Gaussian window function
with $R_G=2$ Mpc. This corresponds  (for the same mass) to an effective
"top-hat" radius of $R(M_G)\approx 2.3$ Mpc/h (as opposed to 1.4 Mpc/h
used in this paper). For the WMAP-3 parameters Weinberg's calculation
would yeild a typicality around $50$\%.

\section{Dependence on simplifying assumptions}
\label{sec:mass}

The calculation of $P(\L)$ that led to the distribution in Fig.~\ref{prho06}
relied on a number of simplifying assumptions. In particular, we
assumed (i) that $\L$ is the only variable parameter and (ii) that the
number of observers is proportional to the mass fraction in giant
galaxies of mass greater than $M_G=10^{12}M_\odot$. We shall now
discuss how sensitive our conclusions are to these assumptions and
what modifications we can expect in more realistic models.

\subsection{Other variable parameters}

A criticism that has often been raised against the anthropic
prediction of $\L$ is that the agreement of the prediction with the
data is destroyed if one allows parameters other than $\L$ to
vary. For example, if the amplitude of the primordial density
perturbations $Q$ is also variable, then galaxies will form earlier in
regions with larger $Q$, and $\L$ can take larger values without
interfering with galaxy formation \cite{Banks}. However, it has been
shown in \cite{Livio} that the probability distribution in this case
factorizes as
\beq
P(Q,\L)dQd\L= P_{prior}(Q)~\erfc\left[{X\over{\sqrt{2}}}\right]dQd\L \propto   
Q^3 P_{prior}(Q)dQ ~\erfc\left[{X\over{\sqrt{2}}}\right]X^2 dX. 
\eeq
Here the factors $Q^3$ and $X^2$ come from the transformation Jacobian
and we have used the expression for $X$ in Eq.~(\ref{X1}).

The value of $Q$ is determined by the shape of the inflaton potential,
and the prior distribution for $Q$ is highly model-dependent
\cite{Hall1,catastrophe,Hall2}. So it is very fortunate that the
$Q$-dependent part of the distribution has completely factored out,
and we are left with the distribution for $X$,
\beq
P(X)dX\propto \erfc\left[{X\over{\sqrt{2}}}\right]d(X^3).
\label{PX}
\eeq
For a fixed $Q$, $X^3\propto\L$ with a constant coefficient, and we
recover Eq.~(\ref{erfc}). For a variable $Q$, Eq.~(\ref{PX}) shows that
the distribution has exactly the same form, except now it should be
regarded as a distribution for the variable $X^3\propto\L/Q^3$. The
typicality of the observed value of $X$ is therefore the same as what
we found for $\L_*$ in the preceding section, ${\cal T}(X_*)=22\%$.

This conclusion can be extended to include a variable dark matter mass
per photon, $\xi$. Using the analysis in \cite{Wilczek}, it can be
shown that, to a good approximation, the distribution in this case can
still be factorized into a part depending on $Q$ and $\xi$ and a part
depending on $X^3\propto\L/Q^3\xi^4$. The predicted range of $X$ is
unaffected by variation of $Q$ and $\xi$ and is independent of their
prior distributions.

\begin{figure}[tb]
\centerline{\epsfxsize=9.0cm\epsffile{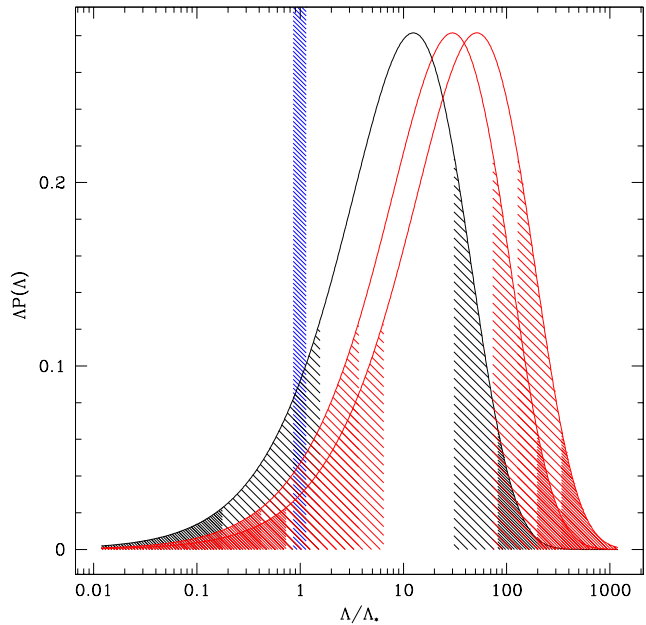}}
\caption{\label{prho_3RG}\footnotesize
The probability distribution for $\Lambda$ assuming three different mass
thresholds for halos that can become habitable galaxies. From left to right, 
the curves correspond to thresholds of $10^{12}M_\odot$,  $10^{11}M_\odot$ and 
$10^{10}M_\odot$ . The typicalities of the observed value, $\L_*$, are $22$\%,
$11$\% and $7$\%, respectively. }
\end{figure}

\subsection{Dependence on the characteristic mass scale $M_G$}

Next we investigate the dependence of $P(\L)$ on the galaxy mass
cutoff $M_G$.  It has been argued by Loeb \cite{Loeb06} that
the anthropic explanation of $\L$ would be significantly weakened if a
large number of habitable planets were found to exist in smaller galaxies, such as
the dwarf descendants of galaxies formed at $z\sim 10$. It would mean
that planet-based observers could be abundant in our Universe even if
the cosmological constant was some orders of magnitude larger than
observed. 

One motivation for introducing the cutoff mass $M_G$ is that, in the
hierarchical structure formation scenario, smaller galaxies typically
form at earlier times and have a higher density of matter. This may
increase the danger of nearby supernova explosions and the rate of
near encounters with stars, large molecular clouds, or clumps of dark
mater \cite{TR98,GV03,Wilczek}. Gravitational perturbations of
planetary systems in such encounters could send a rain of comets from
the Oort-type clouds toward the inner planets, causing mass
extinctions. (Encounters close enough to disrupt planetary orbits are
much less likely.) With the present rate of mass extinctions (about
once in $10^8$ yrs), the intervals between these events are already
comparable to the time it took humans to evolve. Hence, a substantial
increase of the rate may result in a suppression of the number of
observers.

Another consideration is that the properties of galaxies, as a
function of their mass, are observed to undergo a sharp transition at
the halo mass of about $M_h\sim 2\times 10^{11} M_\odot$ (stellar mass
$\sim 3\times 10^{10} M_\odot$) \cite{Kauffmann}. For larger
masses, baryons are efficiently converted into stars, while for
smaller masses the efficiency is lower and decreases as $M$ gets
smaller.  The metallicity is also observed to drop with mass
\cite{Tremonti03}: dwarf galaxies appear to be deficient in heavy
elements, which are necessary for the formation of planets.  These
traits are attributed to supernova feedback: supernova explosions
either expel a fair fraction of gas from low-mass galaxies or inhibit
star formation. (For a discussion and references, see
\cite{Deckel}). Low mass fraction in stars and low metallicity
indicate that dwarf galaxies are less likely to harbor observers.

The effective cutoff mass $M_G$ cannot be determined without a
quantitative study of these effects. We shall not attempt this
here. In order to quantify the dependence of our predictions on the
value of $M_G$, we repeated the analysis in Section III using two lower
values, namely, $10^{11}M_\odot$ and $10^{10}M_\odot$. The resulting
distributions are shown in Fig.~\ref{prho_3RG}.

As expected, a decrease of $M_G$ shifts the distribution towards
larger values of $\L$. However, the effect is not dramatic. The
observed value remains in the 95\% range of the distribution.  Its
typicality ${\cal T}_+$ is reduced from 22\% for $M_G=10^{12}M_\odot$
to $11$\% and $7$\% (for $M_G=10^{11}M_\odot$ and
$M_G=10^{10}M_\odot$, respectively).

\section{Variable $m_\nu$ and joint variation of $m_\nu$ and $\L$}
\label{sec:nu}

\begin{figure}[tb]
\centerline{\epsfxsize=9.0cm\epsffile{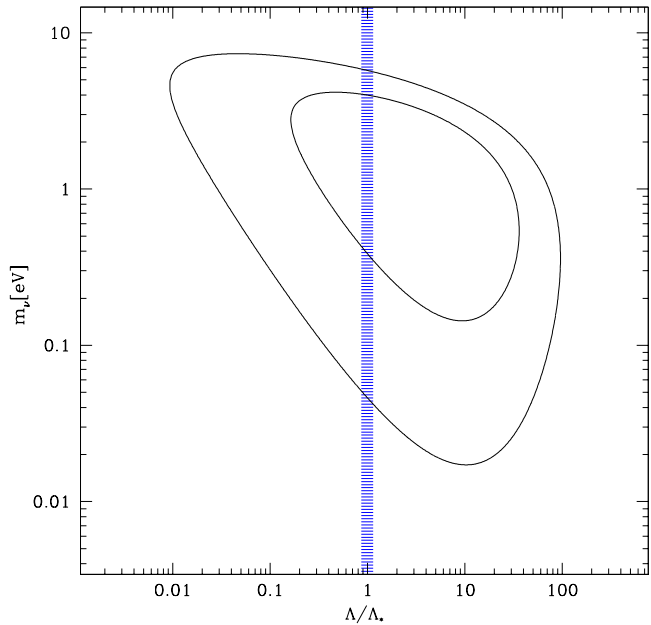}}
\caption{\label{joint_wmap3}\footnotesize
The $68$ and $95$\% joint probability contours for $\Lambda$ and
$m_\nu$ obtained using the WMAP-3 parameters.}
\end{figure}

\begin{figure}[tb]
\centerline{\epsfxsize=9.0cm\epsffile{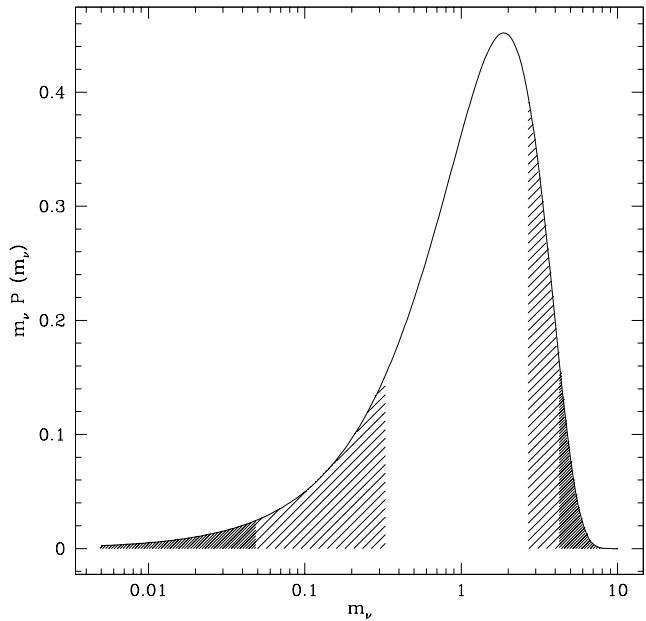}}
\caption{\label{pnu_wmap3}\footnotesize
The probability distribution for the sum of the neutrino masses,
$m_\nu$, obtained using the WMAP-3 parameters.}
\end{figure}

We now consider the case when both the sum of the neutrino masses,
$m_\nu\equiv m_1+m_2+m_3$, and the vacuum energy density, $\L$, vary
from one part of the Universe to another. The possibility that the
smallness of the neutrino masses may be due to anthropic selection has
been first suggested in \cite{TVP03}. The mass-squared differences of
the three neutrino species are now known with an accuracy of about
10\%,
\beq
\Delta m_{32}^2\approx 2.6\times 10^{-3}~eV^2,
\label{m32}
\eeq
\beq
\Delta m_{21}^2\approx 8\times 10^{-5}~eV^2.
\label{m21}
\eeq
The absolute values of the neutrino masses are currently unknown. This
makes them an ideal target for anthropic predictions. 

It is usually assumed that the neutrino mass eigenvalues form a
``normal'' hierarchy, i.e., $m_1\ll m_2\ll m_3$. Then it follows from
Eqs.~(\ref{m32}),(\ref{m21}) that $m_\nu\sim 0.06~eV$. The opposite
regime is when the three masses are degenerate, $m_1\approx m_2\approx
m_3$. An intermediate case is that of inverted hierarchy, $m_3\ll
m_2\approx m_1$, in which case $m_\nu\approx 0.11$ eV.  The
astrophysical upper bound from the WMAP-3 data is $m_\nu\lesssim
2~eV$. Stronger bounds have been quoted from combined astrophysical
data sets, but their validity is not certain (for a recent review see
\cite{Fogli06}).

Assuming that $m_\nu$ and $\L$ vary independently, their joint prior
probability is a product of the individual priors. We have already
chosen a flat prior for $\L$ in Section~\ref{sec:plambda}.  The prior
probability for $m_\nu$ was discussed in \cite{TVP03,PVT04} with the
conclusion that ${\cal P}_{prior}(m_\nu)={\rm const}$ is the best
motivated choice. As in Section~\ref{sec:plambda}, we approximate the
anthropic selection factor as the asymptotic mass fraction in galaxies
of mass $M>M_G$,
\ba
f_{selec}(m_\nu,\L)\propto F(M>M_G,m_\nu,\L) &\propto& 
\erfc\left[{X\over\sqrt{2}}\right],
\label{nGnurho}
\ea
where
\ba
X 
={\delta_c \over \hat \sigma(M)} 
{ {\hat D}(f_{{\nu}*},\L_*,x_*) \over D^{\infty}(f_\nu,\L)} \ .
\label{X2rhonu} 
\ea
The notation above is similar to that of Eq.~(\ref{X2}) of
Section~\ref{sec:plambda}, except for the hats indicating the
quantities evaluated for our local region assuming massless neutrinos.
We refer the reader to \cite{PVT04} for a detailed derivation of the
above expressions. The updated joint probability distribution for $\L$
and $m_\nu$ is shown in Figure~\ref{joint_wmap3}.

It is of interest to also evaluate the anthropic prediction for
$m_\nu$ under the assumption of fixed $\L=\L_*$. This distribution is
shown in Figure~\ref{pnu_wmap3}. The difference, compared to the
results in \cite{TVP03}, is that the mean value of $m_\nu$ has
decreased from $2.1$eV to $1.5$eV, and the lower boundary of the
$2\sigma$ region from $0.07$ eV to $0.05$eV.

Both the single-variable and combined distributions are peaked at a
value $m_\nu\sim 1$~eV, corresponding to degenerate neutrino
masses. The $1\sigma$ range of the distribution, $0.3~eV < m_\nu <
3~eV$, also predicts degeneracy. The prediction, however, is not very
strong: the $2\sigma$ range is (marginally) consistent with the
hierarchical mass pattern. We note that for $m_\nu\sim 1$~eV the
observed value of $\Lambda$ is well within the $1\sigma$ range of the
distribution.

\section{Discussion}
\label{sec:conclude}

The observed value of the cosmological constant, $\L_*$, is mysterious
for two reasons. First, all particle physics estimates yield enormous
values, many orders of magnitude greater than $\L_*$.  Second, $\L_*$
is comparable to the average matter density at the present epoch -- a
coincidence calling for an explanation.  Both of these facts are
accounted for in the multiverse model, where $\L$ is a variable
changing from one place in the universe to another. Despite many
attempts, there are no plausible alternative explanations. A
satisfactory agreement between the observed $\L_*$ and the anthropic
prediction is therefore hard to dismiss.  It may be our first evidence
for the existence of the multiverse.

In this paper we have updated the anthropic probability distribution
for $\L$ using the WMAP-3 data release. This resulted in an improved
agreement between the theory and the data. The typicality of the
observed value $\L_*$, defined as the probability of all values ``less
likely'' than $\L_*$, is ${\cal T}_+(\L_*)=22\%$.  (As we explained in
Sec. II, this estimate should be regarded as a lower bound, since our
analysis did not include negative $\L$.)

We analyzed the dependence of the successful prediction for $\L$ on
the simplifying assumptions used to derive the distribution
$P(\L)$. Our conclusion is that the prediction is rather robust. It is
not drastically modified by variation of the cutoff galactic mass
$M_G$ and remains essentially unaffected by inclusion of other
anthropic variables, such as the amplitude of primordial density
perturbations $Q$, or mass of dark matter per baryon $\xi$. Variation of the
sum of the neutrino masses $m_\nu$ does have some effect on
$P(\L)$. As $m_\nu$ is increased from its lowest value, the agreement
with the data improves, and for $m_\nu\sim 1$~eV the observed value
$\L_*$ is well within the $1\sigma$ range of the distribution.

We have also updated the anthropic prediction for $m_\nu$. With a flat
prior, the distribution is peaked at $m_\nu\sim 1$~eV, suggesting
degenerate neutrino masses.  A hierarchical neutrino mass pattern is
marginally acceptable at the $2\sigma$ level.

Unlike the case of the cosmological constant, there is a viable
alternative explanation for the smallness of the neutrino masses. It
is the see-saw mechanism. We note that a recent study failed
to find models exhibiting this mechanism in a wide class of string
theory inspired models \cite{Langacker}.  We note finally that a value
of $m_\nu\sim 1$~eV is suggested by the Heidelberg-Moscow double beta
decay experiment \cite{doublebeta}. This claim, however, remains
controversial \cite{controversial}.

\acknowledgments

We thank Jaume Garriga, Avi Loeb and Don Page for useful 
discussions. The work of AV is supported in part by grant PHY-0353314
from The National Science Foundation and by grant RFP1-06-028 from The
Foundational Questions Institute.

\end{document}